# Anisotropy of electron momentum distribution in $Bi_2Sr_2CaCu_2O_{8+\delta}$ superconductor studied by positron annihilation technique

Mahuya Chakrabarti[1], A. Sarkar[1,4], S. Chattapadhayay[1], D. Sanyal[2,*], A. K. Pradhan[3], R. Bhattacharya[1] and D. Banerjee[1]

[1]*Department of Physics, University of Calcutta, 92 Acharya Prafulla Chandra Road, Kolkata 700009, India*
[2]*Variable Energy Cyclotron Centre, 1/AF Bidhannagar, Kolkata 700064, India*
[3]*Department of Physics, University of Virginia, McCormick Road, Virginia 22901, USA*
[4]*Department of Physics, Bangabasi Morning College, 19 Rajkumar Chakraborty Sarani, Kolkata 700 009, India*

[*]Corresponding author. Tel.: +91-33-2337-1230; FAX: +91-33-2334-6871
*E-mail address:* dirtha@veccal.ernet.in

**Abstract**

The temperature dependent (30 K to 300 K) Doppler broadening of the positron annihilated γ-radiation (DBPAR) measurement has been carried out on single crystalline $Bi_2Sr_2CaCu_2O_{8+\delta}$ (Bi-2212) high $T_c$ superconducting sample along two different crystallographic orientations. It has been observed that throughout the temperature range the electron momentum distribution has a relatively large value along the crystallographic c-axis than a-b plane. The temperature dependent DBPARL analysis shows a step like increase of *S*-parameter at the temperature region 92 K to116 K.

# 1. Introduction

Positron annihilation technique is a well known nuclear solid-state technique [1] to study the electronic structure, defect properties, electron density distribution and the electron momentum distribution (EMD) in a material. Doppler broadened positron annihilation γ-radiation lineshape (DBPARL) parameters ($S$ and $W$-parameters) are very useful to understand the momentum distribution of electrons in a material. Employing positron annihilation technique on the conventional superconductors like A-15 compounds no measurable changes of the positron annihilation parameters across the superconducting transition have been observed. After the discovery of the high temperature superconductors (HTSC) there have been a large number of efforts [2-8] to study the changes in electron density distribution and electron momentum distribution due to superconducting transition by positron annihilation techniques. It is observed that the changes in the positron annihilation parameters due to the superconducting transition in the HTSC samples are very small, still measurable [8]. Many groups including ours have reported [2-5,7] temperature dependent DBPARL studies in these high temperature superconductors. But the scenario till now is not conclusive.

Our previous work of positron annihilation lifetime measurement [6] on $(Bi_{0.92},Pb_{0.17})_2Sr_{1.91}Ca_{2.03}Cu_{3.06}O_{10+\delta}$ polycrystalline HTSC reports a step like decrease of mean positron lifetime below superconducting transition temperature, $T_c$ and DBPARL experiments at very close interval of temperature for polycrystalline $(Bi_{0.92},Pb_{0.17})_2Sr_{1.91}Ca_{2.03}Cu_{3.06}O_{10+\delta}$ [5] and single crystalline $YBa_2Cu_3O_{7+\delta}$ [7] give an anomalous change in the $S$-parameter with temperature. In the polycrystalline HTSC sample presence of large number of defects as well as grain boundaries may affect the positron annihilation parameters. Considering these facts good quality single crystalline Bi-2212 HTSC has been chosen in the present experiment to explore more reliable results.

In the DBPARL technique positron from the radioactive ($^{22}$Na) source is thermalized inside the material under study and annihilate with an electron emitting two oppositely directed 511 keV γ-rays [1]. Depending upon the momentum of the electron (p) these 511 keV γ-rays are Doppler shifted by an amount $\pm \Delta E$ in the laboratory frame. Where

$$\Delta E = p_L c/2$$

$p_L$ is the component of the electron momentum, $p$, along the detector direction. By using a high resolution HPGe detector one can measure the lineshape of these 511 keV γ-rays.

A widely discussed phenomenon regarding high $T_c$ superconducting oxides is the anisotropy of its properties in different crystallographic direction [9]. The structure of the unit cell of such HTSC materials consists of a stack of conducting (Cu-O plane) and non conducting planes [9]. Crystallographic "c-axis" is the axis perpendicular to these planes. Among the HTSC compounds Bi-2212 is the most anisotropic in nature. Here an effort has been made to observe the manifestation of such anisotropy in the electron momentum distribution by using DBPARL technique.

## 2. Experimental outline

The $Bi_2Sr_2CaCu_2O_{8+\delta}$ crystals were grown by the traveling-solvent floating-zone technique [10]. The high quality cleaved single crystals used in the present DBPARL experiment have the dimension of 3 mm × 5 mm × 0.15 mm with optically smooth surfaces. The superconducting transition temperature $T_c$ of these samples is 91 K [11].

Doppler broadened positron annihilation γ-radiation line-shape (DBPARL) has been measured by an HPGe detector of efficiency 13 %. It has a resolution of 1.10 keV for the 514-keV γ-ray line of $^{85}Sr$ with 6 μs shaping time constant in the spectroscopy amplifier. The two detector coincidence technique [12] has been used to achieve the higher peak to background ratio in the measured $N(E)$ vs. $E$ spectrum under the 511 keV photo-peak. A $2^{//} \times 2^{//}$ NaI(Tl) crystal coupled to a RCA 8850 photomultiplier tube has been placed at 180° with the HPGe detector for the purpose of coincidence measurement. The detection of the oppositely directed 511 keV γ-rays by the NaI detector reduces the background of the annihilation γ-ray spectrum recorded in the HPGe channel under the 511 keV photo-peak and adds to the precision in the direction of the measurement.

About 10 μCi of $^{22}NaCl$ has been deposited directly on the a-b surface of one of the single crystalline $Bi_2Sr_2CaCu_2O_{8+\delta}$ HTSC sample and then cover it with another identical single crystalline $Bi_2Sr_2CaCu_2O_{8+\delta}$ HTSC sample. Area of the deposited source on the surface of the crystal is ~ 1 mm in diameter. Adoption of this procedure completely eliminates annihilation of positron in the source cover and its contribution to the DBPARL spectrum.

Source-sample sandwich has been placed inside a vibration free helium cryogenerator (APD Cryogenics Inc., model number DMX-20) for maintaining the

sample at low temperatures in the range 300K to 30K. We choose a vibration free helium cryogenerator to reduce the possibility of the distortion of the DBPARL spectrum. The system temperature has been controlled by a temperature controller (Scientific Instruments Inc. 9620-1) with ±0.5 K temperature stability. The silicon diode thermometer used in the experiment has been calibrated against a calibrated Pt-resistance thermometer.

For each temperature ~ $2 \times 10^6$ coincidence counts have been recorded under the photo peak of the 511 keV γ-ray at a rate of 110 counts per second. The energy per channel of the multichannel analyzer is kept at 79.6 eV. Background has been calculated from 607 keV to 615 keV energy range of the spectrum. The peak to background ratio is obtained as 14000:1. The system stability has been checked frequently during the progress of the experiment.

The Doppler broadening of the positron annihilated 511 keV γ-ray spectrum has been analyzed by evaluating the so called lineshape parameters [1] (*S*-parameter and *W*-parameter). The *S*-parameter is calculated as the ratio of the counts in the central area of the 511 keV photo peak ( | 511 keV - $E_\gamma$ | ≤ 0.85 keV ) and the total area of the photo peak ( | 511 keV - $E_\gamma$ | ≤ 4.25 keV ). The *S*-parameter represents the fraction of positron annihilating with the lower momentum electrons with respect to the total electrons annihilated. The *W*-parameter represents the relative fraction of the counts in the wings region (1.6 keV ≤ |$E_\gamma$ -511 keV| ≤ 4 keV) of the annihilation line with that under the whole photo peak ( | 511 keV - $E_\gamma$ | ≤ 4.25 keV ). The *W*-parameter corresponds to the positrons annihilating with the higher momentum electrons. The statistical error is 0.2 % on the measured lineshape parameters.

To study the anisotropy of the EMD of the single crystalline Bi-2212 HTSC sample DBPARL experiments have been carried out in two different crystallographic orientations. The first orientation is such that the crystallographic c-axis of the "ordered single crystal" makes an angle $0^o$ (position A) with the joining axis of the HPGe and NaI(Tl) detectors. In this orientation one can probe the c-axial component of the electron momentum as $p_L$ is directed along the c-axis of the single crystal. In the other orientation the angle is $90^o$ (position B) to probe the component of the electron momentum along the a-b plane of the single crystal.

## 3. Results and discussion

The distribution of positrons in these layered-structured HTSC oxides is not uniform. The annihilation characteristics of the positrons from such a structurally complex material bear the information related to the region where positron density distribution is maximum. The positron density distribution calculations for $Bi_2Sr_2CaCu_2O_{8+\delta}$ HTSC by Sundar et al., [4] show that the positron density is maximum in the Bi-O planes. It is attributed that a fraction of positrons are mainly annihilating at the oxygen site [6] of the Bi-O plane. These annihilating 511 keV γ-rays are less Doppler broadened and contribute to the central part of the Doppler broadened 511 keV γ-ray spectrum. Thus the variation of *S*-parameter with temperature may be correlated with the variation of the momentum of electron at the oxygen site of Bi-O plane.

Fig. 1 shows the variation of *S*-parameter with sample temperature for the two different orientations of the single crystalline Bi-2212. As mentioned earlier *S*-parameter reflects the contribution of low momentum electrons in the Doppler broadened 511 keV spectrum whereas the *W*-parameter is associated with the higher momentum electrons. A change in the *S*-parameter is associated with the redistribution of electron momentum inside the material. In the present experiment we observe an increase of *S*-parameter in the temperature region from 92 K to 116K. The observed step like increase of *S*-parameter at 116 K has a magnitude of ~ 0.8 %, which is in agreement with the earlier results [6,8]. Just at the superconducting transition temperature, $T_c$ (91 K) *S*-parameter suddenly comes back nearly to its original value. The typical feature of *S* vs. *T* variation is similar in the two different orientations of the crystal (position A and position B) indicating a common mechanism involved with the superconductivity in all directions. The *S/W*-parameter represents the fraction of lower momentum electron over their higher momentum counterparts. The same behaviour of *S/W*-parameter and *S*-parameter with temperature is visible from figs. 1 & 2. It represents that transfer of electrons [3,13,14] between higher and lower momentum states is associated in the process of superconducting transition which starts far above $T_c$.

The increase of the *S*-parameter suggests either the positrons are less annihilating with the core electrons or an increase of the number of lower momentum electrons at the positron annihilation site. This step-like increase of *S*-parameter above $T_c$ and its coming back to the original value at $T_c$ may be linked with the possibility of the local structural changes [15] which in a way favour the "charge transfer model" valid for these cuprate superconductor [16]. According to the charge transfer model due to the onset of superconductivity charge has been transferred from the Bi-O plane to Cu-O plane (superconducting plane). This could be possible by considering a structural change which helps to increase the coupling of the p-type state in the Bi-O band in such a way that the number of the d-type electrons in Cu-O band decreases. In this way the effective hole density in the Cu-O layer and the electron density at the Bi-O layer (positron annihilation site) increases. In our earlier experiments of temperature dependent positron lifetime studies on Bi-based polycrystalline HTSC [6] a step-like decrease of mean positron lifetime around the superconducting transition region has been observed, which supports the charge transfer model. The increased number of electrons in the Bi-O band increases the probability of positrons to be annihilated with lower momentum electrons of the oxygen site and therefore increases the value of $S$-parameter. Thus the structural changes favour electron momentum redistribution at and around the superconducting transition region.

The most important result is the difference of $S$-parameter value for the two different orientations of the sample (Fig. 1.). The value of $S$-parameter in the entire temperature range (30K to 300K) is higher for position A than position B. The difference between the magnitudes of the $S$-parameter in these two orientations ($\Delta S = S$ at position A $- S$ at position B) is a measure of the anisotropy of the EMD. Fig. 1 shows that $\Delta S$ is almost constant with temperature. Such type of temperature independent anisotropy in the EMD has been observed by probing high $T_c$ superconductors by 2D-ACAR [17] and Compton scattering experiment [18]. The authors of Ref. 18 have observed a significant amount (~ 1 %) of anisotropy in the EMD in the Bi-2212 superconductor. Presently observed anisotropy in the EMD by DBPARL technique is more than 2 % which is twice as observed by the Compton scattering technique. Our results indicate an increased value of electron momentum at the Bi-O plane along the c-axis than a-b plane of the Bi-2212 crystal.

## 4. Conclusion

$S$-parameter vs. temperature graph of the single crystalline $Bi_2Sr_2CaCu_2O_{8+\delta}$ high $T_c$ superconducting sample shows a step like increase in the value of the $S$-parameter (0.8 %) at the temperature region of 116 K (far above $T_c$) to 92 K. Anisotropy in electron momentum distribution between the a-b plane and the c-axis has successfully been probed. This anisotropy has been found to be temperature independent.

## Acknowledgement

One of the authors (S.C) gratefully acknowledges the CSIR, New Delhi for providing financial assistance.

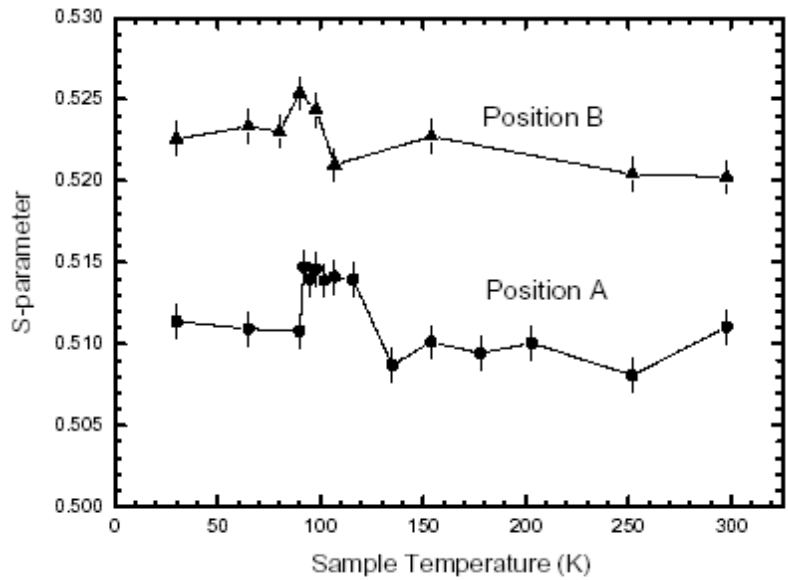

Fig. 1. Variation of *S*-parameter as a function of sample temperature for the two different orientations of the single crystalline Bi-2212 HTSC. For "Position A" crystallographic c-axis of the "ordered single crystal" makes an angle $0°$ with the joining axis of the HPGe and NaI(Tl) detectors. For "position B" the angle is $90°$.

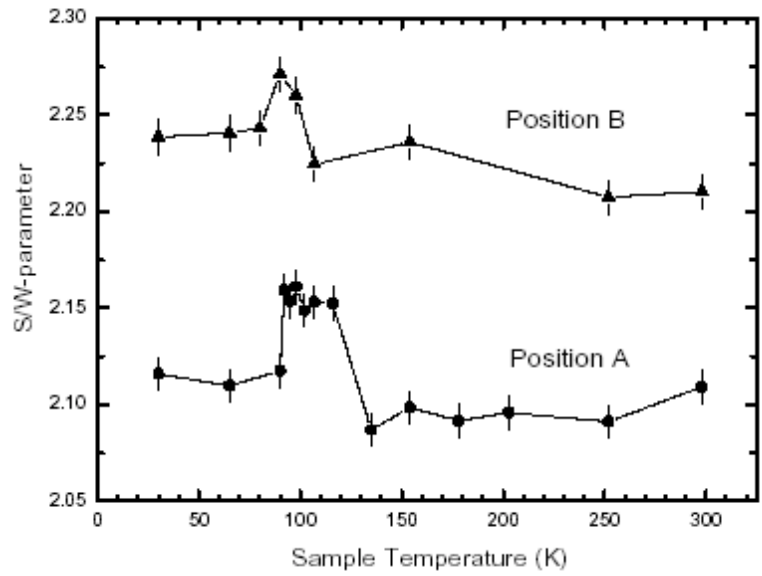

Fig. 2. Variation of *S/W*-parameter as a function of sample temperature for the two different orientations of the single crystalline Bi-2212 HTSC. For "Position A" crystallographic c-axis of the "ordered single crystal" makes an angle $0^o$ with the joining axis of the HPGe and NaI(Tl) detectors. For "position B" the angle is $90^o$.